\documentclass[journal,10pt,twocolumn]{IEEEtran}
\usepackage{graphicx, times, amsmath, amsfonts, comment,amsthm}
\usepackage{amssymb, enumitem, epstopdf, soul, color, cite}
\usepackage[noend]{algorithmic}
\usepackage{algorithm, array}
\usepackage{epstopdf}
\usepackage{subfig}
\usepackage{graphicx}

\newcommand{\beq}{\begin{equation}}
\newcommand{\eeq}{\end{equation}}
\newcommand{\tbf}{\textbf}
\newcommand{\tit}{\textit}

\newcommand{\ud}{\mathrm{d}}

\theoremstyle{plain}
\newtheorem{defcounter}{Definition}
\newtheorem{definition}[defcounter]{Definition}
\theoremstyle{plain}

\theoremstyle{plain}

\theoremstyle{plain}

\theoremstyle{plain}

\theoremstyle{plain}

\newcommand {\Ebb}{\mathbb{E}}

\newcommand {\Vbb}{\mathbb{V}}

\begin{document}
\title{Battery Recharge Time of a Stochastic Linear and Non-Linear Energy Harvesting System}
\author{Sudarshan Guruacharya, Vandana Mittal, and Ekram Hossain\thanks{The authors are with the Department of Electrical and Computer Engineering, University of Manitoba, Canada (email: \{sudarshan.guruacharya, ekram.hossain\}@umanitoba.ca, mittalv@myumanitoba.ca). The work was supported by the Natural Sciences and Engineering Research Council of Canada (NSERC).} 
}
\maketitle

\begin{abstract}
Systems with energy harvesting capability from stochastic sources have been widely studied in the literature. However, the determination of the recharge time of such systems has not received as much attention as it deserves. Here, we examine the recharge time of a battery/super-capacitor when the energy arrival is a discrete stochastic process. We consider the cases when the energy storage system is modeled as a linear and a non-linear system. The energy arrival is assumed to be a Poisson process, or more generally, a renewal process; while the energy packet size may assume any distribution with finite mean and variance. We obtain formulas for the distribution and the expected value of the recharge time. Monte-Carlo simulations verify the obtained formulas.
\end{abstract}


\section{Introduction}
In recent years, ambient energy harvesting and its applications have become a topic to great interest. Basically, a device is assumed to harvest energy from random energy source for its future use \cite[and reference therein]{El-Sayed2016,Ku2016}. 
 From the system theoretical perspective, the harvested energy can be predictable, semi-predictable, or unpredictable. Furthermore, we can also distinguish the harvested energy as being continuous or discrete \cite{Ku2016}. Non-traditional sources of energy such as bodily motion can be modeled as a discrete, stochastic source of energy. Continuous energy can, however, be discretized by suitable sampling method.
 
When we consider an energy harvesting system, the non-ideal behavior of the system arises from the battery's\footnote{By slight abuse of word, by \tit{battery} we refer to any generic energy storage device, such as an electro-chemical battery or a super-capacitor. } self-discharge as well as the inefficiency of the energy conversion process. We say that the system is linear if the efficiency of the energy conversion is fixed. If the efficiency is dependent on the internal state of the system (i.e. the amount of stored energy), then we refer to such systems as non-linear \cite{Gorlatova2013,Biason2016}. Since the self-discharge of a battery is very small, the non-ideal behavior of an energy harvesting system can largely be attributed to the non-linearity of the system. 

While the determination of the state-of-charge of a battery has been extensively studied, only recently some aspects of the recharge time were investigated in \cite{Mishra2015,Mishra2016} where the input power is fixed and continuous, making the system deterministic. The authors obtained the time it takes the system to recharge up to a certain voltage for ideal as well as non-ideal super-capacitor models \cite{Zubieta2000}. However, to the best of our knowledge, recharge time for systems with stochastic, intermittent energy availability is yet to be studied.

In this paper, we will focus on discrete, stochastic energy sources, where energy arrives as a random stream of impulses. Instead of dealing with equivalent circuit models and the related terminal voltages and currents, as done in \cite{Mishra2015,Mishra2016}, we will deal with energy concepts, which allows easier modeling of the system. The accounting of energy is also made easier due to the law of conservation of energy. 

Our contributions in this letter are to (1) model such energy sources as a renewal process and (2) analyze the time it takes to recharge a battery up to a certain fixed level for (3) linear and non-linear storage systems. During the course of our analysis, we will demonstrate that the battery energy of a non-linear system is related to the incoming energy via a logistic function. This result corresponds with empirically determined relationship between input power and harvested power in \cite{Boshkovska2015}. We will also show how the recharge time for non-linear system is related to the recharge time of a linear system. Thus, we will first study the linear system before dealing with non-linear system. We have applied the results of this paper in the study of harvest-then-consume protocol \cite{Guruacharya2017}. {\em We believe that this work can help in further modeling and optimization of energy harvesting systems}.


\section{System Model, Assumptions, and Definitions}
Consider an energy harvesting system where the harvested energy is stored in a battery before being consumed. An \tit{energy outage} is said to occur when the battery becomes empty. A \tit{recharge process} is triggered whenever an energy outage event occurs. During the recharge  process, the battery is initially empty 
and the external energy consumption is turned off. If the system relies only on its energy harvesting capability to recharge itself, then we can model the energy accumulated in the battery, $U(t)$, at any given time $t \geq 0$ after energy outage, as
\beq 
U(t) = \sum_{i=1}^{N_A(t)} \eta_i h(t-t_i; X_i) - \int_0^t p(t)\ud t,   
\label{eqn:recharge-general-model}
\eeq
where $X_i$ is the energy packet size, $p(t)$ is the battery's possibly time varying self-discharge rate, $\eta_i \in (0,1)$ is the recharge efficiency, and $h(t;X)$ is the transient of charging process of the battery given $X$. Here, $h(t; X)$ is any function such that $h(t;X) =0$ for $t<0$ and $\lim_{t \to \infty}h(t; X) = X$. Based on the behavior of $\eta_i$, we will refer to the energy storage system as \tit{linear} if $\eta_i$ is constant and \tit{non-linear} if $\eta_i$ depends on the amount of energy stored \cite{Gorlatova2013}. Lastly, the $N_A(t) = \min\{k: A_0 + A_1 + \cdots + A_k \leq t\}$ is the counting process of the arriving energy, where $A_{i\geq1}$ is the inter-arrival time and $A_0$ is the residual time. If the expression in (\ref{eqn:recharge-general-model}) is negative, then $U(t) =0$.

For example, if energy is delivered to the battery as impulses for a short period of time from $t=0$ to $T$ at a fixed rate $q$, then the total energy transferred is $X = qT$. In wireless power transfer, it is generally assumed that $q$ randomly varies from one impulse to another due to multi-path fading, but is fixed over the short duration $T$. In this scenario, the received power $q$ can be further modeled as $q = \xi d^{-\alpha} \bar{q}$, where $\xi$ is the channel gain, $d$ is the distance between the wireless power transmitter and the energy harvester, $\alpha$ is the path loss exponent, and $\bar{q}$ is the transmit power. The charging process of the battery is given by a piecewise ramp function 
\[ h(t; X) = \left\{ \begin{array}{lcr} 
				qt,  & \mathrm{if} & 0 \leq t \leq T, \\
				X,   & \mathrm{if} & t > T. 
				\end{array} \right. \]
Re-written, we have $h(t;X) \equiv qT g(t) = X g(t)$, where $g(t)$ is given by a normalized version of the above piecewise ramp function
\[ g(t) = \left\{ \begin{array}{lcr} 
				\frac{t}{T}, & \mathrm{if} & 0 \leq t \leq T, \\
				1,  & \mathrm{if} & t > T. 
				\end{array} \right. \]
Thus, a simple manner in which $X$ can modify $g(t)$, an underlying transient  function, is by scaling its amplitude, $h(t; X) \equiv X g(t)$. For very short period $T \to 0$, $g(t)$ can be idealized as a unit step function. We can then neglect the ramp part and simply account for the total energy transferred as $h(t;X) = X$.

For an \tit{ideal system}, we will assume that: (i) there is no self-discharge, $p(t) = 0$; (ii) $h(t; X) = X g(t)$, where $g(t)$ is a unit step function; and (iii) recharge efficiency is linear, $\eta_i = \mathrm{constant}$, which we take as unity. In a \tit{realistic system}, the self-discharge rate of a battery is very small compared to the recharge rate and can be neglected. Likewise, for impulsive energy arrivals, which last for very short durations, assuming $h(t; X)$ to be a step function is also a reasonable approximation. As such, the non-idealness of the system can be largely attributed to the non-linearity of $\eta_i$. Therefore, we will focus our subsequent analysis on linear and non-linear systems without self-discharge. 

For the recharge process of an ideal system, we have a simplification of (\ref{eqn:recharge-general-model})  as 
\beq 
U(t) = \sum_{i=1}^{N_A(t)} X_i.
\label{eqn:recharge}
\eeq 

We now ask for the time required to recharge the battery beyond some desired level. This is essentially a {\em first passage time problem} \cite{Beichelt2002}. Let the first passage time for $U(t)$ given in (\ref{eqn:recharge-general-model}) to cross some level $u > 0$ be $\tau(u) = \inf_t \{t : U(t) > u \}$. Obviously, $u \leq U_{\max}$, where $U_{\max}$ is the maximum battery capacity. Here $\tau$ is a random variable for a given $u$.  For an ideal system, $U(t)$ in (\ref{eqn:recharge}) is a pure jump process; hence we have the events $\{\tau(u) < t\} \equiv \{U(t) > u\}$ to be equivalent. Thus,
\beq
P(\tau(u) \leq t) = P(U(t) > u). 
\label{eqn:level-time}
\eeq

\begin{definition}[Renewal process]
A sequence of arrival times $\{t_n\}$ is a \tbf{renewal process} if $t_n = A_0 + A_1 + \cdots + A_{n-1}$, where the inter-arrival times $\{A_{i\geq1}\}$ given by $A_i = t_{i+1} - t_i$ are mutually independent, non-negative random variables with common distribution $F_A$ such that $F_A(0) = 0$. Here $A_0$ is the time for the first arrival and is known as \tbf{residual time}. If there is an arrival at the origin, that is  $A_0 = 0$, then renewal process is said to be a \tbf{pure} renewal process. Otherwise, the renewal process is said to be a \tbf{delayed} renewal process.
\label{def:renewal-process}
\end{definition}

We can also represent $\tau(u)$ by the decomposition
\beq 
\tau(u) = A_0 + \sum_{i=1}^{N_X(u)} A_i,
\label{eqn:tau-decomposition}
\eeq
where $N_X(u) = \min\{k: X_1 + \cdots + X_k \leq u\}$ is again a counting process with $X_i$ as the energy packet size.

Let the mean and variance of $A$ be finite. Also, $\{X_i\}$ is assumed to be a sequence of non-negative random variables, with common distribution $F_X$ such that the mean and variance are finite. We assume that $\{A_i\}$ and $\{X_i\}$ are independent of each other. Lastly, we assume that the random vectors $\{(A_i, X_i) \}$ are identically distributed as $(A,X)$. For notational convenience, we will denote $\lambda = 1/\Ebb[A]$ and $\bar{X} = \Ebb[X]$.

Note that the renewal process becomes a Poisson process when the inter-arrival times are exponentially distributed. The Poisson process can also arise as a result of superposition of common renewal processes \cite{Cox1954}. Thus, Poisson energy arrival can be used to model situations where multiple harvesters send energy, according to a common renewal process, to a common battery. Likewise, when the inter-arrival time is deterministic, the renewal process can model the slotted time models. Hence, the renewal process is a generalization of these special cases.


\section{Analysis of Linear Storage System}
The $N_A(t)$ in (\ref{eqn:recharge}) will be a pure renewal process only when there is an energy arrival immediately after the energy outage event, which triggers the recharge process. Since this is an unrealistic expectation, $N_A(t)$ is more properly modeled as a \tit{delayed renewal process}. In this case, the epoch of $n$-th energy arrival $t_n$ is given by $t_n = A_0 + A_1 + \cdots + A_n$, where the random variable $A_0$ has a different distribution from the inter-arrival times $A_i$, $i \geq 1$. Only for Poisson process, due to its memoryless property, $A_0 \stackrel{d}{=} A_i \stackrel{d}{=} \mathrm{Exp}(\lambda)$. 

Assuming that the energy outage event, which triggers the recharge process, occurs a long time after the system has been in operation\footnote{By this, we mean that the energy arrival process begins at $t=-\infty$, while the recharging process begins at $t=0$.}, the residual time $A_0$ converges in distribution to  $f_{A_0}(t) = \frac{1 - F_A(t)}{\mu_A}$, where $f_{A_0}$ is the density of $A_0$, $F_A$ is the inter-arrival distribution, and $\mu_A$ is the mean of $A$. Thus, the resulting renewal process after the energy outage event becomes an \tit{equilibrium (or stationary) renewal process} (see \cite[Theo. 4.1]{Beichelt2002}). 
In the following subsections, we will first investigate the special case when the energy arrival follows a Poisson process, and then later, more generally, for the case when the energy arrival follows a renewal process.


\subsection{When Energy Arrival Follows a Poisson Process}
If energy arrival follows a Poisson process, then the distribution of $U(t)$ in (\ref{eqn:recharge}) can be obtained by conditioning on $N_A(t) = n$, and using total probability theorem as
\[ P(U(t) \leq u) = e^{-\lambda t} \sum_{n=0}^{\infty} \frac{(\lambda t)^{n}}{n!} F_X^{(n)}(u), \]
where $F_X^{(n)}$ is the $n$-fold convolution of $F_X$ defined recursively as $F_X^{(i)}(x) = \int F_X^{(i-1)}(x - t) \; \ud F(t)$ where $i = 1,2,3, \ldots, n$; and  $F_X^{(0)}(x)$ is a unit step function at the origin. Thus, from (\ref{eqn:level-time}) the distribution of the level crossing time is 
\beq 
P(\tau(u) < t) = 1 - e^{-\lambda t} \sum_{n=0}^{\infty} \frac{(\lambda t)^{n}}{n!} F_X^{(n)}(u). 
\label{eqn:general-poisson}
\eeq

Except for few packet size distributions (like exponential, see \tbf{Appendix}), it is difficult to evaluate (\ref{eqn:general-poisson}) exactly for general $F_X$ distribution. However, if we make normal approximation for $F_X^{(n)}$, which is the distribution for the sum of $n$ independent random variables, as per the central limit theorem, as $F_X^{(n)}(x) \approx \Phi(\frac{x - n\bar{X}}{\sigma_X \sqrt{n}})$, where $\Phi(\cdot)$ is the CDF of standard normal distribution, we have
\beq 
P(\tau(u) < t) = 1 - e^{-\lambda t} \sum_{n=0}^{\infty} \frac{(\lambda t)^{n}}{n!} \Phi \left(\frac{u - n\bar{X}}{\sigma_X \sqrt{n}} \right). 
\label{eqn:approx-general-poisson}
\eeq

Since $\tau$ is a non-negative random variable, the $k$-th moment of recharge time can then be obtained as $\Ebb[\tau^k] = \int_0^\infty k t^{k-1} P(\tau \geq t) \ud t$.  Using this identity, the expected recharge time is 
\begin{align}
\Ebb[\tau(u)] &= \int_0^\infty P(\tau(u) \geq t) \; \ud t \nonumber \\
&= \sum_{n=0}^{\infty} \Phi \left(\frac{u - n\bar{X}}{\sigma_X \sqrt{n}} \right) \cdot \frac{1}{n!}  \int_0^\infty (\lambda t)^{n}  e^{-\lambda t} \; \ud t  \nonumber \\
&= \frac{1}{\lambda} \sum_{n=0}^{\infty} \Phi \left(\frac{u - n\bar{X}}{\sigma_X \sqrt{n}} \right), \label{eqn:expected-tau-approx-general-poisson}
\end{align}
where the last step is because  $n! = \lambda \int_0^\infty (\lambda t)^{n} e^{-\lambda t} \; \ud t$.


\begin{figure*}[t]
\begin{center}
\subfloat[]{\includegraphics[width=\columnwidth]{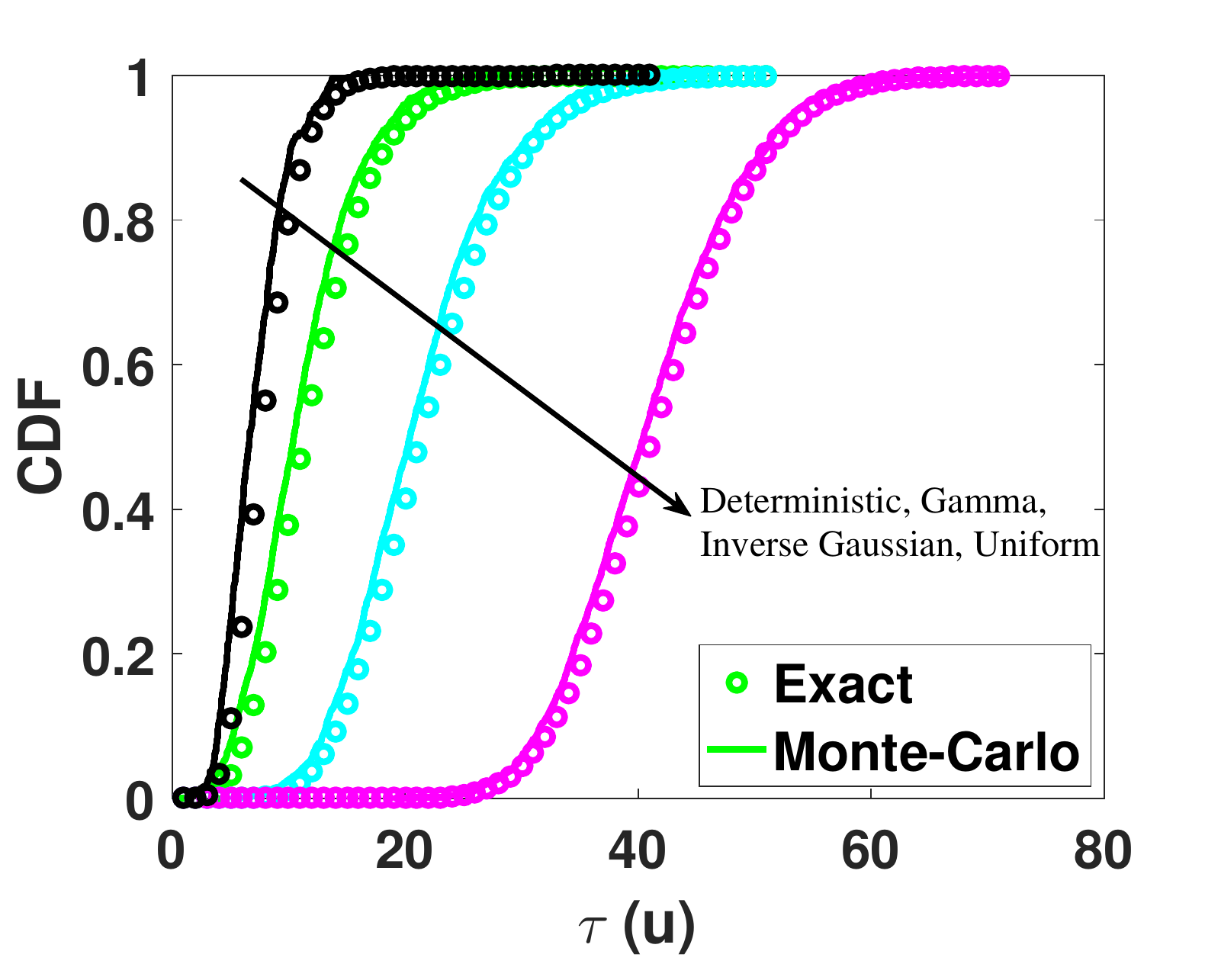}
\label{fig:LinearPoisson}
}\hspace{0.001in} 
\renewcommand{\thesubfigure}{b}
\subfloat[]{\includegraphics[width=\columnwidth]{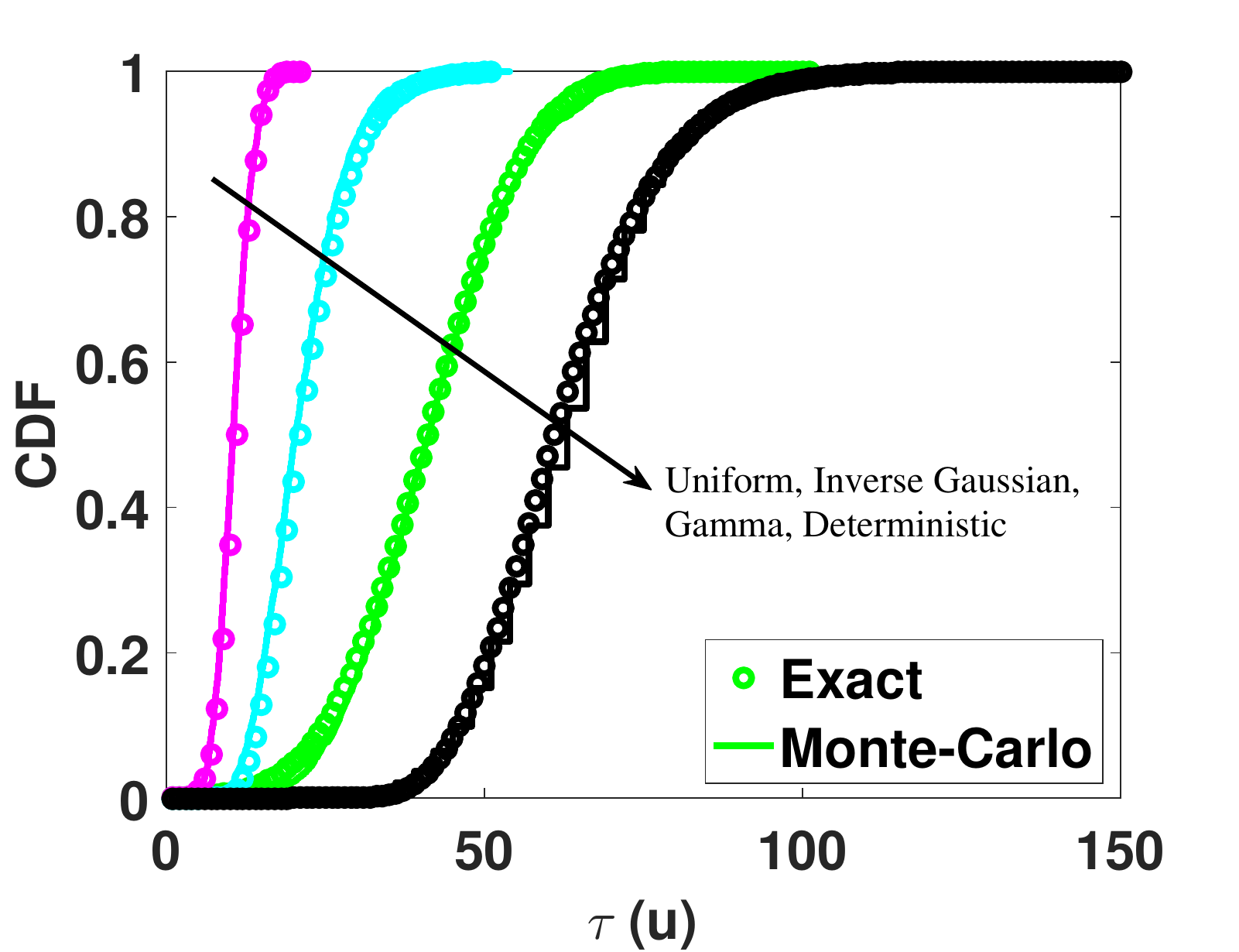}
\label{fig:LinearRenewal}%
}
\caption{CDF of recharge time for linear system with $u = 20$ (a) When energy arrival is a Poisson process and packet sizes are deterministic, gamma, inverse Gaussian, and uniform distributed. (b) When energy arrival is a renewal process where packet sizes are exponentially distributed and energy arrivals are deterministic, gamma, inverse Gaussian, and uniform distributed. }
\label{fig:LinearSystem}
\end{center}
\end{figure*}

\begin{figure*}[t]
\begin{center}
\subfloat[]{\includegraphics[width=\columnwidth]{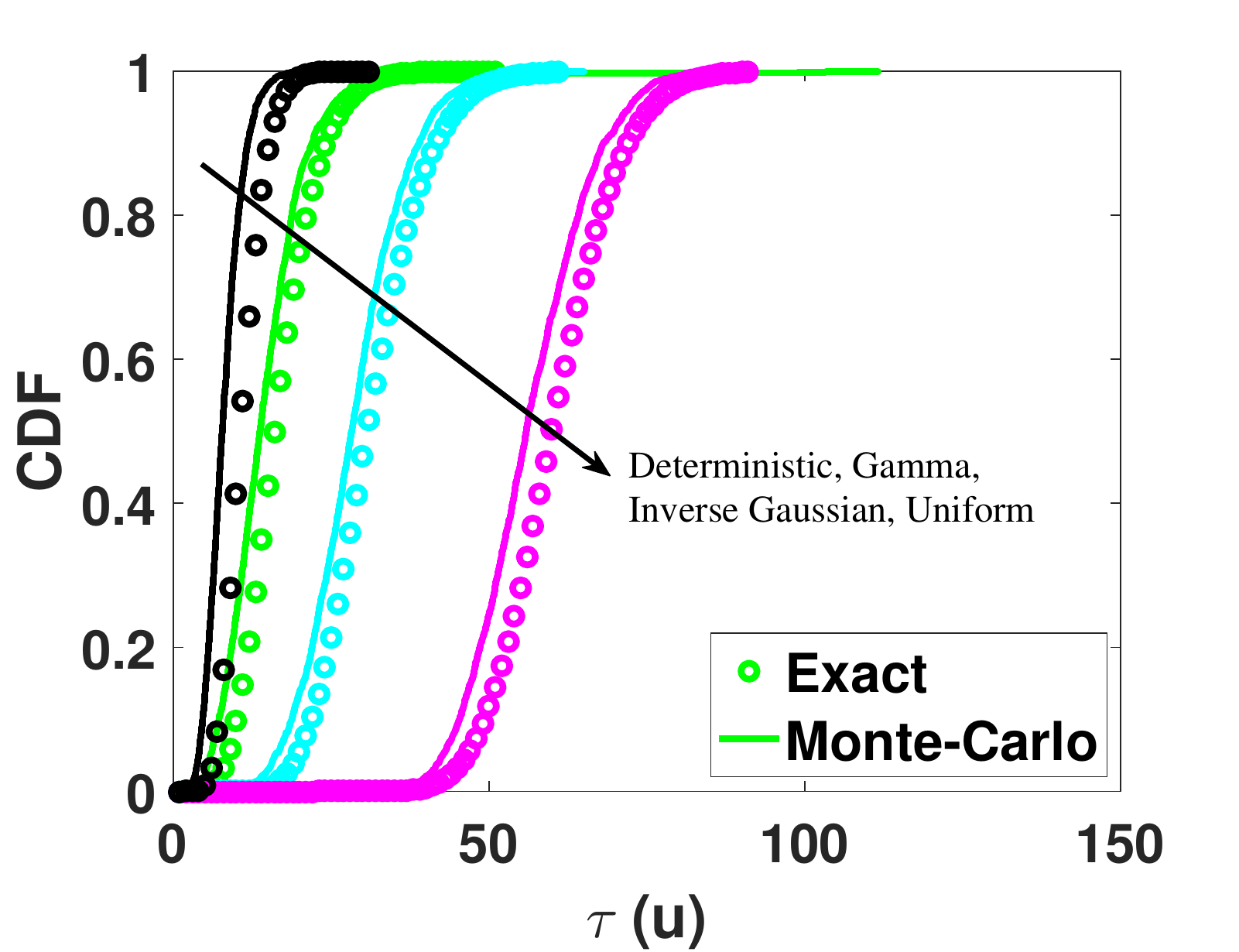}
\label{fig:non-linearPoisson}
}\hspace{0.001in} 
\renewcommand{\thesubfigure}{b}
\subfloat[]{\includegraphics[width=\columnwidth]{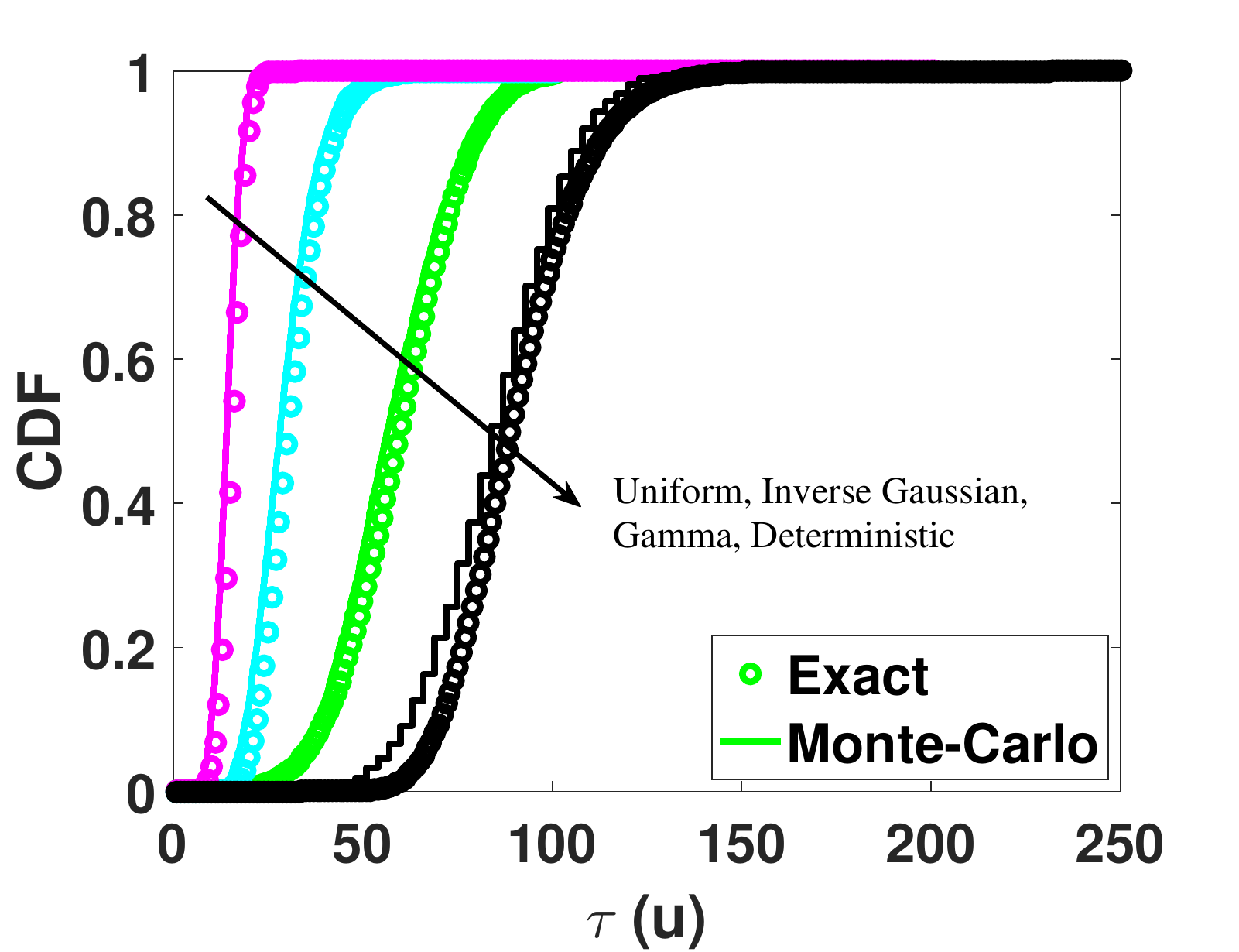}
\label{fig:non-linearRenewal}%
}
\caption{CDF of recharge time for non-linear system with $u$ = $20$, $U_{\max}$ = $25$, $\beta$ = $1.1$ (a) When energy arrival is a Poisson process and packet sizes are deterministic, gamma, inverse Gaussian, and uniform distributed. (b) When energy arrival is a renewal process where packet sizes are exponentially distributed and energy arrivals are deterministic, gamma, inverse Gaussian, and uniform distributed. }
\label{fig:non-linearSystem}
\end{center}
\end{figure*}


\subsection{When Energy Arrival Follows a Renewal Process}

In full generality, since both $\{A_i\}$ and $\{X_i\}$ defines a renewal process, here we will directly work with (\ref{eqn:tau-decomposition}), where $A_0$ is differently distributed from $A_i$, $i\geq1$. The sum $\sum_{i=1}^{N_X(u)} A_i$ is in itself a pure renewal process, while $\tau(u)$ is a stationary, delayed renewal process. 

By the linearity of expectation and Wald's identity, \[ \Ebb[\tau(u)] = \Ebb[A_0] + \Ebb[N_X(u)]\Ebb[A].\] Here for the stationary renewal process, the expected value of the residual time $A_0$ is $\Ebb[A_0] = \frac{\mu_A^2 + \sigma_A^2}{2\mu_A} = \frac{1 + \lambda^2 \sigma_A^2}{2\lambda}$.
Also, asymptotically\footnote{Here $f(x)\sim g(x)$ if and only if $\lim_{x\to\infty} \frac{f(x)}{g(x)} = 1$.} for pure renewal process \cite[Theo. 4.5]{Beichelt2002}\cite{Smith1959} $\Ebb[N_X(u)] \sim \frac{1}{2}(\bar{X}^{-2} \sigma_X^2-1) + \bar{X}^{-1} u$. Hence, asymptotically 
$\Ebb[N_X(u)]\Ebb[A] \sim \frac{1}{2 \lambda}(\bar{X}^{-2} \sigma_X^2 - 1)+ (\lambda \bar{X})^{-1} u$. Therefore, putting everything together and after some basic algebra, we have the mean value of $\tau(u)$  asymptotically as, 
\beq 
\Ebb[\tau(u)] \sim  \frac{\lambda \gamma^2}{2 \bar{X}^2} + \frac{u}{\lambda \bar{X}},
\label{eqn:expected-tau-renewal}
\eeq
where $\gamma^2 = ( \lambda^{-2}\sigma_X^2  + \sigma_A^2 \bar{X}^2)$. For large $u$, the constant term may be neglected, and we obtain $\Ebb[\tau(u)] \sim (\lambda \bar{X})^{-1} u$.


Similarly, we can analyze the variance of $\tau(u)$ using the total variance theorem as
\begin{align*} 
\Vbb[\tau(u)] &= \Vbb[A_0] + \Ebb[\Vbb[\tau|N_X]] + \Vbb[\Ebb[\tau|N_X]] \\
&= \Vbb[A_0] + \Ebb[N_X \Vbb[A]] + \Vbb[N_X \Ebb[A]]  \\
&= \Vbb[A_0] + \Ebb[N_X] \Vbb[A] + \Vbb[N_X] (\Ebb[A])^2.
\end{align*}
For the pure renewal process \cite{Smith1959}, 
$\lim_{u\to\infty} \frac{1}{u} \Vbb[N_X(u)] = \bar{X}^{-3} \sigma_X^2$, so we have the asymptotic $\Vbb[N_X(u)] \sim \bar{X}^{-3} \sigma_X^2 u$. Similarly, the variance of residual time is 
$\Vbb[A_0] = \frac{\mu_A^{(3)}}{3\mu_A} - \left(\frac{\mu_A^2 + \sigma^2_A}{2 \mu_A}\right)^2$, where $\mu_A^{(3)}$ is the third moment of $A$. Note that $\Vbb[A_0]$ is a constant and is independent of $u$. Putting everything together and after some basic algebra, we have the asymptotic for $\Vbb[\tau(u)]$ as
\begin{align}
\Vbb[\tau(u)] &\sim 
\Vbb[A_0] + \frac{\gamma^2 u}{\bar{X}^3}, 
\label{eqn:variance-tau-renewal}
\end{align}
where $\gamma^2 = ( \lambda^{-2}\sigma_X^2  + \sigma_A^2 \bar{X}^2)$. For large value of $u$, we can neglect the first constant term, so that $\Vbb[\tau(u)] \sim \gamma^2 \bar{X}^{-3} u$.

From the central limit theorem, we have for large $u$

\beq
P(\tau(u) \leq t) \approx \Phi\left(  \frac{t - (\lambda \bar{X})^{-1} u}{\gamma \bar{X}^{-3/2} \sqrt{u} } \right). 
\label{eqn:normal-approx-renewal}
\eeq
Greater accuracy can be obtained by including the neglected constant terms in the mean and variance in (\ref{eqn:normal-approx-renewal}).


\section{Analysis of Non-Linear Storage System}
In a non-linear energy storage system, the efficiency of the charging process is dependent on the amount of energy stored in the system. Here we still assume that $p(t) = 0$ and $h(t)$ is a step function in (\ref{eqn:recharge-general-model}). In the model suggested by \cite{Gorlatova2013, Biason2016}, we have the non-linear efficiency given by
\beq 
\eta(U) = 1 - \left(\frac{U-a}{b}\right)^2, 
\label{eqn:non-linear-efficiency}
\eeq
where $a = \frac{1}{2}U_{\max}$ and $b = \beta (\frac{U_{\max}}{2} ) $. Here, $U_{\max}$ is the battery capacity and $\beta > 1$ is the non-linearity parameter. Note that $\eta \to 1$ as $\beta \to \infty$. 

To find the instantaneous relationship between the input energy $X$ and the stored energy $U$, we solve the differential equation $\frac{\ud U}{\ud X} = \eta(U)$ with initial condition $U = 0$ when $X=0$.  Hence, we have the integral $\int \ud X  = \int \frac{\ud U}{\eta(U)}$, whose solution is given by the arc hyperbolic tangent function
\beq 
 X = b \tanh^{-1} \left( \frac{U - a}{b} \right) + C,
 \label{eqn:input-energy-non-linear}
\eeq
 where $C$ is the constant of integration. Using the initial condition, we obtain $C = b \tanh^{-1} \left( \frac{a}{b} \right)$. Substituting the expressions for $a$ and $b$, we get $C = \frac{\beta U_{\max}}{2} \tanh^{-1} \left(\frac{1}{\beta}\right)$. Rearranging (\ref{eqn:input-energy-non-linear}), we have  
\beq 
U = a + b\tanh \left(\frac{X - C}{b}\right). 
\label{eqn:stored-energy-non-linear}
\eeq

Equation (\ref{eqn:stored-energy-non-linear}) tells us how to transform a linear system into a non-linear system. A similar logistic relation was empirically proposed in \cite{Boshkovska2015}. Thus, given the aggregate harvested energy, $\sum_{i=1}^{N_A(t)} X_i$, we can now find the distribution of the stored energy as
\[ P(U(t) \leq u) = P \left(\sum_{i=1}^{N_A(t)} X_i \leq C+b\tanh^{-1} \left(\frac{u-a}{b}\right)\right). \]

We see that effect of non-linearity is equivalent to changing the threshold energy level of the linear system. Since we have assumed no self-discharge, the $U(t)$ for non-linear system is also a pure jump process. Thus, from (\ref{eqn:level-time}) we have 
\beq 
P(\tau(u) \leq t) = P(\tau_\ell(u') \leq t), 
\label{eqn:tau-non-linear}
\eeq 
where $\tau_\ell$ is the level-crossing time for linear system and $u' = C+b\tanh^{-1} \left(\frac{u-a}{b}\right)$. For the general renewal process, using (\ref{eqn:normal-approx-renewal}) for the linear system, we have the equivalent formula for non-linear system as
\beq
P(\tau(u) \leq t) \approx \Phi\left(  \frac{t - (\lambda \bar{X})^{-1} u'}{\gamma \bar{X}^{-3/2} \sqrt{u'} } \right). 
\label{eqn:normal-approx-renewal-non-linear}
\eeq


\section{Numerical Verification}

Fig.~\ref{fig:LinearSystem} and Fig.~\ref{fig:non-linearSystem} plots the cumulative distribution function (CDF) for the linear and non-linear storage systems, respectively.  In Fig.~\ref{fig:LinearPoisson} and Fig.~\ref{fig:non-linearPoisson}, we assume energy arrival is a Poisson process; while in  Fig.~\ref{fig:LinearRenewal} and Fig.~\ref{fig:non-linearRenewal}, the energy arrival is a renewal process. The theoretical expressions for first passage time $\tau(u)$ for the Poisson energy arrival is given by (\ref{eqn:approx-general-poisson}) while for renewal process is given by (\ref{eqn:normal-approx-renewal}). In the evaluation of (\ref{eqn:approx-general-poisson}), we truncate after $n=100$. We consider $u = 20$ energy-units and $U_{\max} = 25$ energy-units. For a given distribution and a fixed value of $u$, 2000 simulations were run to construct the empirical CDF of the first passage time.

In Fig.~\ref{fig:LinearPoisson}, while the energy packet size can have any distribution with finite mean and variance,  we consider the case when the energy packet sizes are given by uniform, deterministic, inverse Gaussian, and gamma distributions.  
Here the inter-arrival time $A_i$ is exponentially distributed $\mathrm{Exp}(1)$ and packet sizes $X_i$ are uniform, $\mathrm{U}(0,1)$, deterministic  $\delta(x-3)$, inverse Gaussian, $\mathrm{IG}(1,2)$ and gamma, $\mathrm{Gamma}(1,2)$, distributed. For the more general case, when the energy arrival is a renewal process, there can be any distribution for energy inter-arrival time and packet size. In Fig.~\ref{fig:LinearRenewal},  we plot the CDF for the cases when the packet size is exponentially distributed while the energy inter-arrival times have uniform, deterministic, inverse Gaussian and gamma distributions, parametrized as before. The results from the simulations match closely with the theoretical prediction given by (\ref{eqn:approx-general-poisson}) and (\ref{eqn:normal-approx-renewal}). 

Fig.~\ref{fig:non-linearSystem} shows a more realistic scenario where the system is non-linear. During Monte-Carlo simulations, the battery energy was updated as $U_{i+1} = \eta_i X_{i+1} + U_i$, while the non-linear efficiency $\eta_i$ was updated based on $U_i$ for the new energy packet arrival using (\ref{eqn:non-linear-efficiency}). For theoretical analysis, expression given by (\ref{eqn:normal-approx-renewal-non-linear}) is used. In the evaluation, the value of the non-linearity parameter is taken to be $\beta$ = $1.1$. The results for non-linear system, when energy arrival follows a Poisson process and a renewal process, are shown in Fig.~\ref{fig:non-linearPoisson} and Fig.~\ref{fig:non-linearRenewal}, respectively, parameterized as in the linear case. Here too the results from the simulations match closely with the theoretical prediction.


\section{Conclusion}
We have studied the time it takes for a battery to recharge up to a given level, when the energy source is discrete stochastic. We have examined the cases when the energy arrival is a Poisson process, and more generally, a renewal process, for both the linear as well as non-linear charging processes. Formulas for the distributions of the recharge time and the expected recharge time have been obtained, which have been verified via Monte Carlo simulations.


\appendix

\subsection{Exponential Packet Size}
Given the Poisson arrival, if we further assume that $X \stackrel{d}{=} \mathrm{Exp}(1/\bar{X})$ is exponentially distributed with mean $\bar{X}$, then the $n$-fold convolution of $X$ results in Erlang distribution
\[ F_{U|n}(u) = 1 - e^{-u/\bar{X}} \sum_{i=0}^{n-1} \frac{(u/\bar{X})^i}{i!}, \]
where $n \geq 1$. For $n=0$, $F_{U|n}(u) = 1$. Therefore, we have 
\begin{align*} 
& 1 - P(\tau(u) < t) \\
=\;& e^{-\lambda t} +  e^{-\lambda t} \sum_{n=1}^{\infty} \frac{(\lambda t)^{n}}{n!} \left(1 - e^{-u/\bar{X}} \sum_{i=0}^{n-1} \frac{(u/\bar{X})^i}{i!} \right). 
\end{align*}
Here $e^{-\lambda t} \sum_{n=0}^{\infty} \frac{(\lambda t)^{n}}{n!} = 1$, as per the normalization of Poisson distribution. Thus, the expression simplifies to 
\[ P(\tau(u) < t) = e^{-(\lambda t + u/\bar{X})} \sum_{n=1}^{\infty} \frac{(\lambda t)^{n}}{n!} \sum_{i=0}^{n-1} \frac{(u/\bar{X})^i}{i!}. \]

We can express the truncated exponential sum in terms of Gamma function, as 
$\sum_{i=0}^{n-1} \frac{(u/\bar{X})^i}{i!} =  e ^{u/\bar{X}} \frac{\Gamma(n,u/\bar{X})}{\Gamma(n)}$, where $\Gamma(a,x)$ is the upper incomplete Gamma function. Thus we arrive at the expression for the distribution of $\tau$ as
\beq 
P(\tau(u) < t) = e^{-\lambda t} \sum_{n=1}^{\infty} \frac{(\lambda t)^{n}}{n!} \frac{\Gamma(n,u/\bar{X})}{\Gamma(n)}. 
\label{eqn:exp-poisson}
\eeq

\bibliographystyle{IEEE} 

\begin{thebibliography}{1}

\bibitem{El-Sayed2016} 
A.R. El-Sayed, {\em et al.}, ``A survey on recent energy harvesting mechanisms,'' in Proc. {\em 2016 IEEE Canadian Conf. on Electrical \& Computer Eng. (CCECE'16)}, 
pp. 1--5, 2016.

\bibitem{Ku2016}
M.-L. Ku, {\em et al.}, ``Advances in energy harvesting communications: Past, present, and future challenges,'' {\em IEEE Commun. Surveys \& Tutorials}, no. 2, vol. 18, pp. 1384--1412, 2016.



\bibitem{Gorlatova2013}
M. Gorlatova, A. Wallwater, and G. Zussman, ``Networking low-power energy harvesting devices: Measurements and algorithms,'' {\em IEEE Trans. Mobile Comput.}, vol. 12, no. 9, pp. 1853--1865, Sept. 2013. 

\bibitem{Biason2016}
A. Biason and M. Zorzi, ``On the effects of battery imperfections in an energy harvesting device,” {\em 2016 Int. Conf. Comput. Netw. Commun. (ICNC)}, pp. 1--7, Feb. 2016. 



\bibitem{Mishra2015}
D. Mishra, S. De, and K. R. Chowdhury, ``Charging time characterization for wireless RF energy transfer," {\em IEEE Trans. Circuits Syst. II, Exp. Briefs}, vol. 62, no. 4, pp. 362--366, Apr. 2015.

\bibitem{Mishra2016}
D. Mishra and S. De, ``Effects of practical rechargeability constraints on perpetual RF harvesting sensor network operation," {\em IEEE Access}, vol. 4, pp. 750--765, 2016.

\bibitem{Zubieta2000}
L. Zubieta and R. Bonert, ``Characterization of double-layer capacitors for power electronics applications,'' {\em IEEE Trans. Ind. Appl.}, vol. 36, no. 1, pp. 199--205, Jan./Feb. 2000. 

\bibitem{Boshkovska2015}
E. Boshkovska, {\em et. al.}, ``Practical Non-Linear Energy Harvesting Model and Resource Allocation for SWIPT Systems,'' {\em IEEE Commun. Lett.}, vol. 19, no. 12, pp. 2082 -- 2085, Dec. 2015.

\bibitem{Guruacharya2017}
S. Guruacharya, V. Mittal, and E. Hossain, ``Level-triggered harvest-then-consume protocol with two bits or less energy state information,'' accepted in {\em IEEE Wireless Commun. Lett.}, 2017. Available: [Online]. https://arxiv.org/abs/1709.06928





\bibitem{Beichelt2002}
F.E. Beichelt and L.P. Fatti, {\em Stochastic Processes and Their Applications}. CRC Press, 2002.


\bibitem{Cox1954}
D.R. Cox and W.L. Smith, ``On the superposition of renewal processes,'' {\em Biometrika}, vol. 41, no. 1/2, pp. 91--99, Jun. 1954.

\bibitem{Smith1959}
W.L. Smith, ``On the cumulants of renewal processes,'' {\em Biometrika}, vol. 46, no. 1/2, pp. 1--29, Jun., 1959.


\end{thebibliography}

\end{document}